\pdfoutput=1
\documentclass[aps,prd,a4paper,showpacs,preprintnumbers,floatfix,nofootinbib,twocolumn]{revtex4-1}
\usepackage{graphicx}
\usepackage{url}
\usepackage{xcolor}
\usepackage{amsmath}
\usepackage{amssymb}
\usepackage[toc,page]{appendix}
\colorlet{shadecolor}{gray!15}
\definecolor{greenLinks}{rgb}{0, 0.6, 0} 
\definecolor{blueLinks}{rgb}{0, 0, 0.6}
\definecolor{redLinks}{rgb}{0.6, 0, 0}
\definecolor{tempText}{rgb}{0.55, 0.10,0.67}
\definecolor{eprintLinks}{rgb}{0.4, 0.4, 0.4}
\definecolor{journalLinks}{rgb}{0.6, 0, 0}

\usepackage[T1]{fontenc} 

\usepackage{hyperref}
\hypersetup{colorlinks=true,linkcolor=redLinks,citecolor=greenLinks,
urlcolor=redLinks,
pdfborder={0 0 1}}
\hypersetup{
    bookmarks=true,         
    unicode=false,          
    pdftoolbar=true,        
    pdfmenubar=true,        
    pdffitwindow=false,     
    pdfstartview={FitH},    
    pdftitle={My title},    
    pdfauthor={Author},     
    pdfsubject={Subject},   
    pdfcreator={Creator},   
    pdfproducer={Producer}, 
    pdfkeywords={keyword1, key2, key3}, 
    pdfnewwindow=true,      
    colorlinks=false,       
    linkcolor=red,          
    citecolor=green,        
    filecolor=magenta,      
    urlcolor=cyan           
}

\def\21{$\mathrm{SU(2)_L \otimes U(1)_Y}$ }

\newcommand{\AddrAHEP}{AHEP Group, Institut de F\'{i}sica Corpuscular --
  C.S.I.C./Universitat de Val\`{e}ncia, Parc Cientific de Paterna.\\
  C/Catedratico Jos\'e Beltr\'an, 2 E-46980 Paterna (Val\`{e}ncia) - SPAIN}

\begin{document}

\preprint{IP/BBSR/2017-5}

\title{ 
Resolving the atmospheric octant by an improved measurement of the reactor angle
	}

\author{Sabya Sachi Chatterjee~$^{1,2}$}\email{sabya@iopb.res.in}

\author{Pedro Pasquini~$^3$}\email{pasquini@ifi.unicamp.br}
\author{J.W.F. Valle~$^4$}\email{valle@ific.uv.es, URL:  http://astroparticles.es/}

\affiliation{
$^1$~Institute of Physics, Sachivalaya Marg, Sainik  School Post, Bhubaneswar 751005, India\\
$^2$ Homi Bhabha National Institute, Training School Complex, Anushakti Nagar, Mumbai 400085, India\\
$^3$ Instituto de F\'isica Gleb Wataghin - UNICAMP, {13083-859}, Campinas SP, Brazil\\
$^4$~\AddrAHEP}

\begin{abstract}

  Taking into account the current global information on neutrino
  oscillation parameters we forecast the capabilities of future long
  baseline experiments such as DUNE and T2HK in settling the
  atmospheric octant puzzle.
  We find that a good measurement of the reactor angle $\theta_{13}$
  plays a key role in fixing the octant of the atmospheric angle
  $\theta_{23}$ with such future accelerator neutrino studies.

\end{abstract}

\pacs{12.15.-y,13.15.+g,14.60.Pq,14.60.St} 

\maketitle
\section{Introduction}

The discovery of neutrino oscillations as a result of solar and
atmospheric studies constitutes a major milestone in astroparticle
physics~\cite{Kajita:2016cak,McDonald:2016ixn}.
Earthbound experiments based at reactors and accelerators have not
only provided a confirmation of the oscillation picture but also
brought the field into the precision age.
Despite the great experimental effort, however, two of the oscillation
parameters remain poorly determined, namely the atmospheric mixing
angle $\theta_{23}$ and the CP violating phase
$\delta_{CP}$~\cite{Forero:2014bxa,Capozzi:2016rtj,Esteban:2016qun}.

Underpinning these parameters as well as determining the neutrino mass
ordering constitute important challenges in the agenda of upcoming
oscillation experiments, needed to establish the three-neutrino
paradigm. Concerning the two poorly known neutrino parameters,
$\theta_{23}$ yields two degenerate
solutions~\cite{Forero:2014bxa,Capozzi:2016rtj,Esteban:2016qun},
well known as the octant problem~\cite{Fogli:1996pv}. One of them is
known as lower octant (LO): $\theta_{23}\,< \, \pi/4$ while the other
is termed as higher octant (HO): $\theta_{23}\,> \, \pi/4$.
The role of $\theta_{13}$ and its precise determination has been
stressed in early papers~\cite{Minakata:2002jv,Maltoni:2004ei}.
The actual discovery of large $\theta_{13}$ that has opened a
tremendous opportunity for the long-baseline neutrino oscillation
experiments to resolve the octant issue within the standard 3-flavor
framework. This may however be just an approximation to the true
scenario, which may involve new physics such as
non-unitarity~\cite{valle:1987gv,Miranda:2016ptb,Miranda:2016wdr,Escrihuela:2016ube}
which may have important effects on the propagation of astrophysical
neutrinos~\cite{nunokawa:1996tg,grasso:1998tt} non-standard
interactions~\cite{Wolfenstein:1977ue} as well as a light sterile
neutrino~\cite{Kosmas:2017zbh}.

Recently there have been many papers addressing the octant issue
within the standard 3$\nu$
scenario~\cite{Agarwalla:2013ju,Agarwalla:2013hma,Chatterjee:2013qus,Bass:2013vcg,Bora:2014zwa,Das:2014fja,Nath:2015kjg}.
However it has also been shown recently that the octant sensitivity
may completely change in the presence of the above non-standard
features, i.e.  non-unitarity~\cite{Dutta:2016eks}, non-standard
interaction~\cite{Agarwalla:2016fkh} or a light sterile
neutrino~\cite{Agarwalla:2016xlg}.
  
  In this letter we specifically focus on the reactor angle
  $\theta_{13}$ and on whether an improved precision in its
  measurement from reactors, combined with results from future long
  baseline experiments such as DUNE and/or T2HK, could provide a final
  resolution to the octant puzzle.
  Taking into account current global neutrino oscillation parameter
  fits, we forecast the potential of DUNE~\cite{Acciarri:2015uup} and
  T2HK~\cite{Abe:2015zbg} for pinning down the correct octant of
  $\theta_{23}$.
  We find that a sufficiently good measurement of the reactor angle
  $\theta_{13}$ directly fixes values of $\theta_{23}$ for which the
  octant of the atmospheric angle can be distinguished.

\section{Theory preliminaries}
\label{sec:theory-preliminaries}

Following \cite{Akhmedov:2004ny}, the appearance and survival
oscillation probabilities in the presence of matter can be written
approximately as
\begin{flalign}
\label{eq:oscil1}
P_{\mu e}\,\approx \,& 4 s_{13}^2 s_{23}^2 \sin^2\Delta_{31}  \nonumber \\
& + 2\alpha \Delta_{31} s_{13} \sin 2\theta_{12}\sin 2\theta_{23} \cos(\Delta_{31} \pm \delta_{CP})  \nonumber \\
 =\,& P_0 + P_I  
\\ \label{eq:oscil2}
P_{\mu \mu} \,\approx \,& 1- \sin^22\theta_{23}\sin^2\Delta_{31}-4s_{13}^2 s_{23}^2 \frac{\sin^2(A-1)\Delta_{31}}{(A-1)^2}
\end{flalign}
where $s_{ij} = \sin\theta_{ij}$,
  $\alpha \,=\, \frac{\Delta m_{21}^2}{\Delta m_{31}^2}$,
  $\Delta_{31} \,= \,\frac{\Delta m_{31}^2 L}{4E}$ and the function A
  = $\frac{2EV_{CC}}{\Delta m_{31}^2}$ describes the matter
  profile. Here $V_{CC}$ is the charged current potential in earth
matter, while L and E are the propagation distance and energy of the
neutrinos, respectively. The $\pm$ sign in front of $\delta_{\rm CP}$
corresponds to neutrinos (upper sign) and antineutrinos (lower
sign). The term $P_0$ is the octant sensitive term, whereas the term
related to $\sin^22\theta_{23}$ generates the octant degeneracy.

    An experiment is octant sensitive, if there is always a finite
    difference between the two probabilities corresponding to the two
    octants, despite the minimization performed over the different
    oscillation parameters. Mathematically,
\begin{equation}
 \Delta P \equiv P_{\mu e}^{\rm HO} - P_{\mu e}^{\rm LO} \neq 0
\end{equation}
Note that we assume that one of the two octants is the true octant in
order to generate the data, while the other one is the false octant in
order to simulate the theoretical model predictions. We will always
assume that $\theta_{13}$ lies in its true value
($\sin^2\theta_{13} = 0.0234$) in the true octant. Following
Eq.~\ref{eq:oscil1}, we can write
\begin{eqnarray}
 \Delta P = \Delta P_0 + \Delta P_I~. 
\end{eqnarray}
Now, by expanding Eq.~\ref{eq:oscil1} around $\theta_{23}=\pi/4 \pm \eta$
and $\sin^2\theta_{13}=(1 + \epsilon)\sin^2\theta_{13}$ we get,
\begin{eqnarray}\nonumber
  P_{0}\, = \,(1 \pm 2\eta + \epsilon)P^{'} + O(\epsilon \eta)
\end{eqnarray}
where $\epsilon = \pm \delta(\sin^2\theta_{13}$) denotes the error on
$\sin^2\theta_{13}$ and the $\pm$ sign in front of $\eta$ refers to HO
(upper sign) and LO (lower sign) and,
\begin{eqnarray}
 P^{'} \equiv P^{'}(\theta_{23}=\pi/4,\theta_{13}=\theta_{13}^{true}) = 2s_{13}^2\sin^2\Delta_{31}.\nonumber
\end{eqnarray}
leading to
 \begin{eqnarray}
 \label{eq:oscil3}
 \Delta P_0 = (P_0^{\rm HO} - P_0^{\rm LO}) = P^{'}(4\eta \pm \epsilon)~.
\end{eqnarray}
The double sign in front of $\epsilon$ refers to the $\rm LO^{\rm true}$
(upper sign) and $\rm HO^{\rm true}$ (lower sign). 

In the same manner, we can also write 
\begin{flalign}
\label{eq:oscil4}
 &\Delta P_I  =B \biggl[\sin\theta_{13}^{\rm HO} \cos(\Delta_{31} \pm \delta_{\rm CP}^{\rm HO}) \nonumber \\ &\hspace{3cm}-\sin\theta_{13}^{\rm LO} \cos(\Delta_{31} \pm \delta_{\rm CP}^{\rm LO})\biggr]
\end{flalign}
where, B =
$4\sin\theta_{12}\cos\theta_{12}(\alpha \Delta) \sin\Delta_{31}$.
Notice that, as mentioned above, $\sin\theta_{13}^{\rm HO}$ and
$\sin\theta_{13}^{\rm LO}$ change shape depending on true versus
wrong octant.

For the time being suppose one neglects the error on
$\sin^2\theta_{13}$ by taking $\epsilon \to 0$. $\Delta P_0$ is
positive definite, while $\Delta P_I$ can have either sign due to
the presence of the unknown $\delta_{\rm CP}$. As a result $\Delta P$
may become zero for the unfavorable combinations of octant and
$\delta_{CP}$, so that octant sensitivity can be completely
lost. However, it has been noticed in the literature
\cite{Agarwalla:2013ju,Machado:2013kya} that this type of degeneracy
can be lifted by using both neutrino and antineutrino channels and one
can achieve good octant sensitivity in the 3-flavor scenario.

In the presence of a nonzero error on $\sin^2\theta_{13}$, then
$\Delta P_0$ is also a positive definite quantity, since the current
error on $\sin^2\theta_{23}$ is bigger than the error on
$\sin^2\theta_{13}$, i.e., we can safely assume $\eta > \epsilon$.
In this case it is clear from Eqs.~\ref{eq:oscil3} and \ref{eq:oscil4}
that the unfavorable contribution coming from $\epsilon \neq 0$
changes the magnitude of $\Delta P_0$ and $\Delta P_I$ in such a way
that overall value of $\Delta P$ decreases further than in the
previous case. As a result the octant discrimination sensitivity
decreases significantly even in the presence of neutrino and
antineutrino channels. The larger the error, the less will be the
resulting octant sensitivity. This will be clearly seen in the next
section.

\section{Simulation Details}
\label{sec:simulation-details}Here we present in some detail the
    experimental configurations of the DUNE and T2HK experiments used
    for this work. For a more detailed discussion
    see~\cite{Chatterjee:2017xkb}

    {\bf \textit {DUNE}:} Deep Underground Neutrino Experiment (DUNE)
    is a long-baseline (1300 km) accelerator-based experiment with
    neutrinos travelling from Fermilab to South Dakota. Following the
    DUNE CDR~\cite{Acciarri:2015uup}, we are using a 40 Kt LArTPC as
    its far detector, and a 80 GeV proton beam with beam power 1.07
    MW. A total 300 Kt.MW.yrs of exposure has been assumed with
    neutrino mode running for 3.5 yrs, and the antineutrino mode
    running for 3.5 yrs. All the signal and background event numbers
    have been matched following Table~3.5 and Table~3.6 of
    \cite{Acciarri:2015uup}. With this all the reconstructed neutrino
    and antineutrino energy spectra and sensitivity results have been
    reproduced as close as possible to the same reference. 
    As a simplified case for all the neutrino and antineutrino
    appearance and disappearance channels we have assumed an
    uncorrelated 4\% signal normalization error and 10\% background
    normalization error.

    {\bf \textit {T2HK}:} T2HK (Tokai to Hyper-Kamiokande) is an
    off-axis accelerator based experiment with baseline 295
    km. According to \cite{Abe:2015zbg}, it plans to use the same 30
    GeV proton beam as T2K, provided by the J-PARC facility and a 560
    Kton (fiducial volume) Water Chernkov far detector. An integrated
    beam of power $7.5 \,\rm{MW}\times 10^7$ sec has been assumed for
    this work which corresponds to $1.56\times 10^{22}$ protons on
    target. In order to make the expected event numbers nearly the same for
    neutrinos and antineutrinos, we consider a 2.5 yrs of neutrino
    running mode and 7.5 yrs of antineutrino running mode. All the
    signal and background event numbers have been matched following
    Table~7 and Table~8 of Ref.~\cite{Abe:2015zbg} and all other
    sensitivity results have been reproduced with good agreement. As a
    simplified case we have assumed an uncorrelated 5\% signal
    normalization error, and 10\% background normalization error with
    no energy calibration error.
      
\begin{figure}[h!]
\centering
\includegraphics[width=0.5\textwidth]{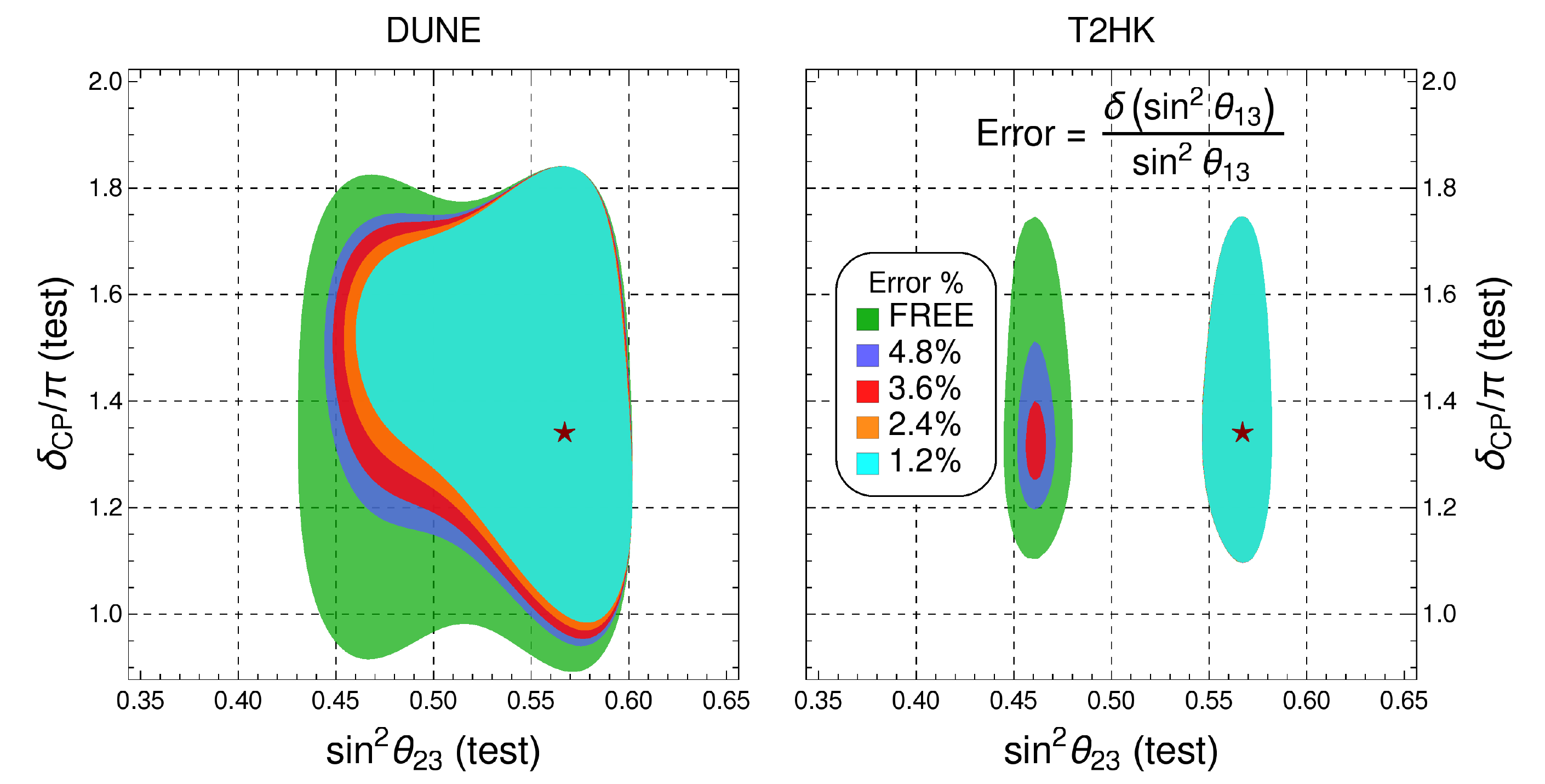}
\includegraphics[width=0.5\textwidth]{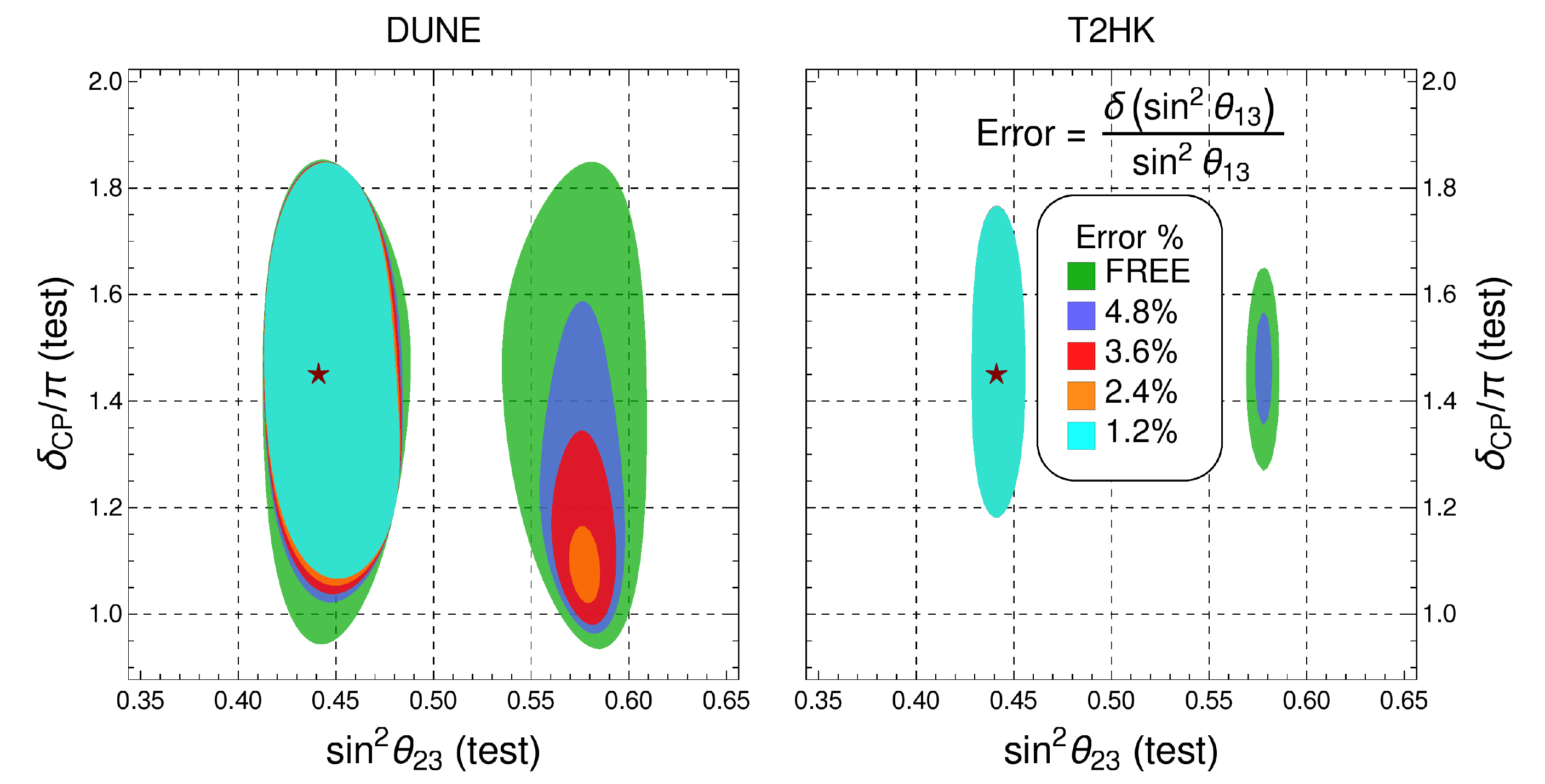}
\caption{\label{fig:12} Precision measurement of $\theta_{23}$ and
  $\delta_{CP}$ at 4$\sigma$ ($\Delta\chi^2$ = 19.33) confidence. Left
  (Right) panels correspond to DUNE (T2HK). Differently shaded
  (colored) regions correspond to various errors associated with
  $\sin^2\theta_{13}$. The Top panel uses the Global Fit
  in~\cite{Forero:2014bxa}, while the Bottom panel corresponds to the
  Global Fit in~\cite{Esteban:2016qun}. The star denotes the bestfit
  point.}
\end{figure}

    We have performed a realistic simulation by using the GLoBES
    package \cite{Huber:2004ka,Huber:2007ji}. The best fit values of
    the oscillation parameters were taken from \cite{Esteban:2016qun}
    except for the top plot of Fig.~\ref{fig:12}, in which we have
    followed \cite{Forero:2014bxa}. They are the following:
  $\sin^2\theta_{12}$ = 0.306, $\sin^2\theta_{13}$ = 0.0216,\,
  $\sin^2\theta_{23}$ = 0.441 for NH ,\, $\delta_{CP}$ = 1.45$\pi$,\,
  $\Delta m_{21}^2$ = 7.5$\times10^{-5}\;\rm eV^2$,\,\,\,and
  $\Delta m_{31}^2$ = 2.524$\times10^{-3}$
  (-2.514$\times10^{-3}$)\;$\rm eV^2$ for NH (IH). Here NH (IH) is
  short for normal hierarchy (inverted hierarchy). In all of our
  numerical analysis, we have assumed NH as fixed both in data and
  theory. In order to determine the sensitivity towards the
  measurement of the octant of $\theta_{23}$, we have defined the
  $\chi^2$ function as,
\begin{equation}
\chi^2=\chi^2_{\rm GLoBES}+\chi^2_{\rm Priors}
\end{equation}
where $\chi^2_{\rm GLoBES}$ is the standard GLoBES Poissionian
chi-squared, while $\chi^2_{\rm Priors}$ is given by,
\begin{equation}\label{eq:prior}
\chi^2_{\rm Priors}=\sum_{i=2,3}\left(\frac{s^{2,\,\rm TRUE}_{1i}-s^{2,\,\rm TEST}_{1i}}{\delta\left(s^{2,\,\rm TRUE}_{1i}\right)}\right)^2
\end{equation}
with $\delta\left(s_{ij}^{2,\,\rm TRUE}\right)$ is the $\sin^2\theta_{ij}^{\rm TRUE}$ error from~\cite{Esteban:2016qun},
while $s^{2,\,\rm A}_{ij}=\sin^2\theta_{ij}^{\rm A}$. Here $A=$ TRUE, TEST
denote the true and test values of the angles respectively. We have
not included either $\delta_{\rm CP}$ or $\theta_{23}$ priors, as we
are focusing on the capability of each experiment to measure them.  In
order to distinguish the true octant from the false one. We define the
chi-squared difference as
$\Delta \chi^2_{\rm oct}=|\chi^2_{\rm min}(\theta_{23}\leq
\pi/4)-\chi^2_{\rm min}(\theta_{23}>\pi/4)|.$
Here $\chi^2_{\rm min}(\theta_{23})$ is the $\chi^2$ function
minimized over other oscillation parameters. Note that one can assume
one of the octants (say, $\theta_{23}\leq \pi/4$) as true and the other
one as false, and vice-versa.

Recent reactor experiments have reached a precision at the percent
level for the measurement of the reactor angle, fixing its central
value around $\sin^2\theta_{13}\sim 0.02$. Current and foreseen
precision levels on the reactor angle are given in
table~\ref{tab:dt13}. For the simulation we took the central value of
$\sin^2\theta_{13}$ from the global fit~\cite{Esteban:2016qun}, and vary the error on $\sin^2\theta_{13}$ as a prior as in Eq.~\ref{eq:prior}. This makes the analysis more
robust, as taking a single experiment error cannot give a general picture for the value of $\sin^2\theta_{13}$.
\begin{table}[h!]
  \centering
  \begin{tabular}{ccccc}
  \hline \hline 
      & DC~\cite{moriond} & RENO~\cite{RENO:2015ksa} & Daya-Bay~\cite{An:2016ses} & Global~\cite{Esteban:2016qun}\\ \hline \hline
  $s_{13}^2/10^{-2}$&  2.85 & 2.09 &  2.09  & 2.34\\
 $\delta\left(s_{13}^2\right)/s_{13}^2$ & 16.7\% & 13.4\% & 4.9\%  & 3.5\% \\
$\delta\left(s_{13}^{2,\rm Expec}\right)/s_{13}^2$ & 10\% & 5\%  & 3.6\%  & <3\%\\ \hline
  \end{tabular}
  \caption{ Current and expected values of the reactor mixing angle
 $\theta_{13}$ and its sensitivity for different experiments and current 
global neutrino oscillation fit. The expected values are based on~\cite{Seo:2017tjb}. The \% is calculated by taking the $1\sigma$ region from the central value. }
    \label{tab:dt13}
\end{table}

\section{Results and discussion}
\label{sec:results-discussions}

In this section we present our numerical results and their
comprehensive discussion. In our whole analysis we have used a
line-averaged constant matter density of 2.95 gm/$\rm cm^3$ for DUNE
and 2.8 gm/$\rm cm^3$ for T2HK within the PREM
\cite{DZIEWONSKI1981297,stacey:1977} profile.

\subsection{Precision measurement}
\label{sec:prec-meas}

Fig.~\ref{fig:12} represents the 4$\sigma$ confidence level
measurement of $\theta_{23}$ and $\delta_{\rm CP}$ for various
combinations of the relative error associated with
$\sin^2\theta_{13}$. The symbol "star" in the body of the plot
corresponds to the best fit value for two assumptions: (I) Top, the
global fit in~\cite{Forero:2014bxa} and (II) Bottom, the global fit
in~\cite{Esteban:2016qun}. The left (right) panel is for DUNE
(T2HK). The cyan band corresponds to 1.2\%, the orange band
corresponds to 2.4\%, the red band is for 3.6\% and the blue band
corresponds to 4.8\% error on $\sin^2\theta_{13}$. In contrast, the
green band is generated by the free marginalization over
$\sin^2\theta_{13}$.
We have marginalized over $\Delta m_{31}^2$ and $\theta_{12}$ with
1$\sigma$ prior on $\sin^2\theta_{12}$ taken from
\cite{Esteban:2016qun}. Left panel, clearly shows that DUNE can not
exclude the wrong octant for errors above $\sim$2.0\%, while it can
surely exclude the wrong octant at 4$\sigma$ confidence if
$\sin^2\theta_{13}$ is more tightly constrained as for the case of
option (II), as seen from the cyan shaded region. Thanks to its higher
statistics in the disappearance channel, T2HK performs better, and can
exclude the wrong octant up to 2.4\% $\sin^2\theta_{13}$ error for the
option (I), and 3.6\% in the case of the option (II), i.e., T2HK can
measure the atmospheric mixing angle very precisely.
\begin{figure}[h!]
 \centering
 \includegraphics[width=0.5\textwidth]{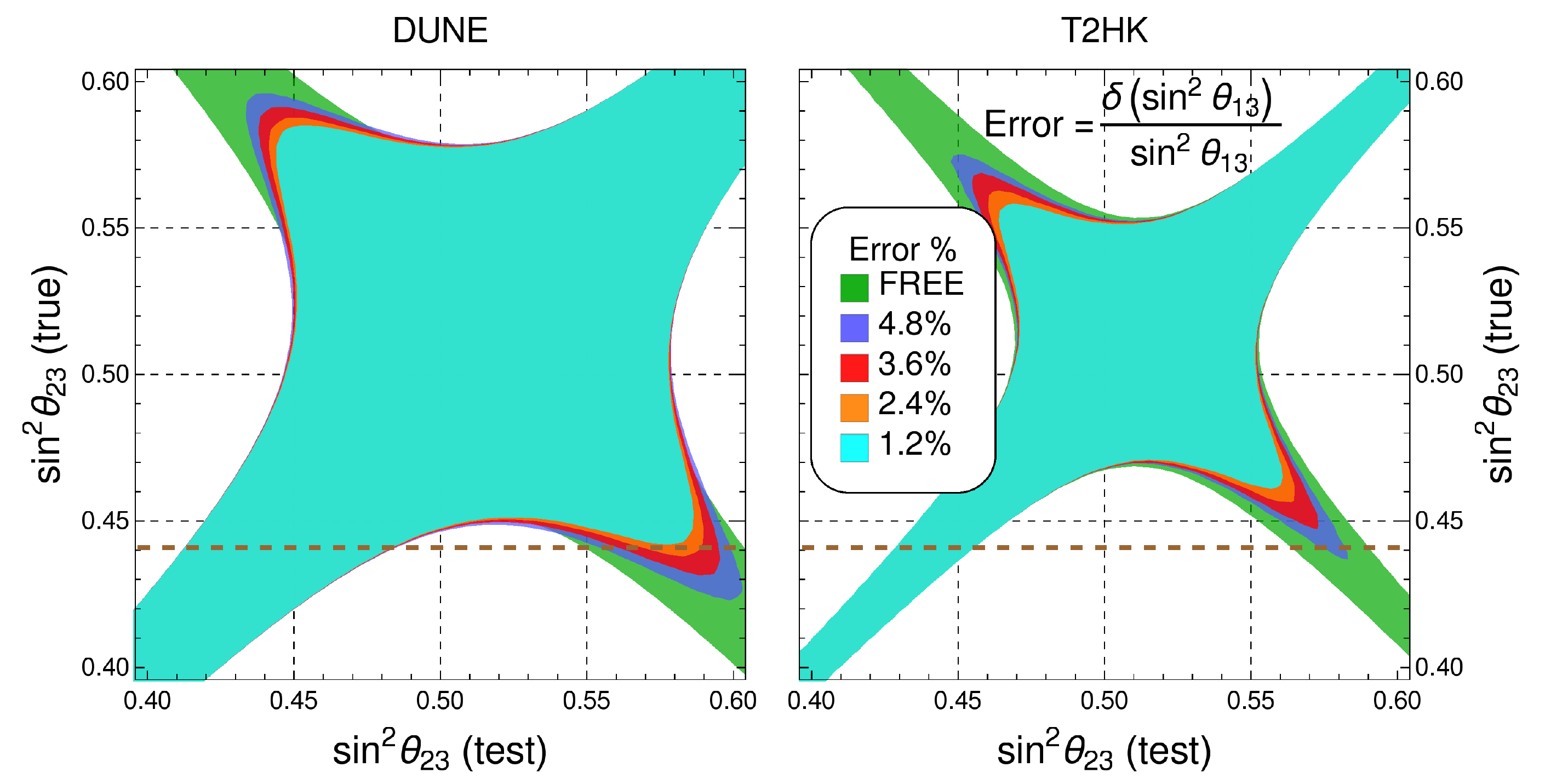}
 \caption{\label{fig:12_all}$4\sigma$ ($\Delta\chi^2$ = 19.33) precision measurement of
   $\theta_{23}$. The left (right) panel is for DUNE
   (T2HK). Differently shaded regions correspond to various errors
   associated with $\sin^2\theta_{13}$. The thick dashed line
   represents the current best fit value from \cite{Forero:2014bxa}.}
 \end{figure}
  
 Note that from Fig.~\ref{fig:12} one can compare the result for two
 oscillation fits.  Although Ref.~\cite{Esteban:2016qun} contains the
 most recent data from Daya-Bay, T2K, and NO$\nu$A, which constrain
 $\theta_{13}$, they do not include the atmospheric data sample, as
 in~\cite{Forero:2014bxa}. While the latter has an impact upon which
 is the preferred octant, it has worse precision on the
 $\sin^2\theta_{13}$, which plays a role in the octant
 discrimination. The two analyses are therefore complementary, though
 an update of~\cite{Forero:2014bxa} is clearly desirable (work in this
 direction is currently underway).
 Our work shows that, in both cases, DUNE by itself will not be enough
 to solve the octant problem, while T2HK can have a better chance to
 uncover the right value of the atmospheric angle.
 In contrast to the determination of the neutrino mass ordering,
 despite the unprecedented level of precision on $\theta_{13}$ that
 may be reached in future studies, there will always be a region that
 is octant blind in any experiment, close to the maximality limit.  
 
 We now turn to a very general question, namely, how well can these
 two experiments measure $\theta_{23}$ irrespective of its true value
 chosen by nature.

Fig.~\ref{fig:12_all} addresses this issue. The
 simulation procedure is exactly the same as for fig.~\ref{fig:12},
 except for the fact that we have marginalized over $\delta_{\rm CP}$
 both in the data and the theory. As a result this figure represents
 the most conservative scenario. If we draw a horizontal line for each
 true value of $\sin^2\theta_{23}$, it touches the different colored
 shaded regions associated to different $\sin^2\theta_{13}$
 errors. The horizontal boundary of each touched shaded region
 corresponding to a particular color represents the 4$\sigma$
 uncertainty on $\sin^2\theta_{23}$. It can be determined simply by
 looking at the brown thick dashed line at $\sin^2\theta_{23}$(true) =
 0.441 and focusing on the cyan band. This procedure extracts all the
 relevant information coming from fig.~\ref{fig:12}. It is noticeable
 that DUNE measures the LO ($\sin^2\theta_{23}$(true)\,$<$ 0.45)
 better than the HO. However, the performance of T2HK is substantially
 higher than that of DUNE in both the octants. An important
 consequence of the green area in Fig.~\ref{fig:12_all} is the fact
 that neither DUNE nor T2HK can distinguish the octant without prior
 knowledge of $\theta_{13}$.
\begin{figure}[h!]
\centering
\includegraphics[width=0.5\textwidth]{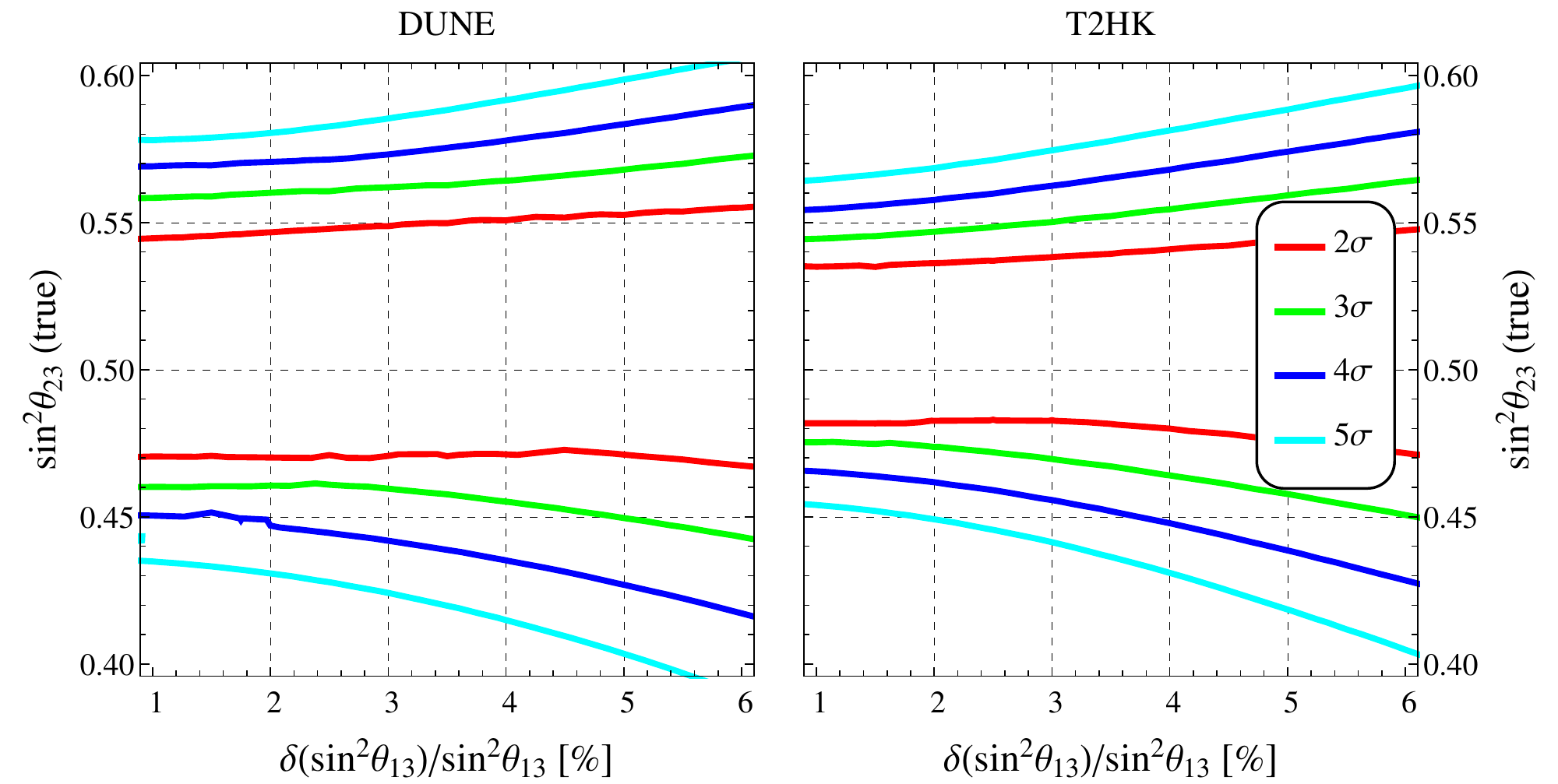}
\caption{\label{fig:13} Octant discrimination potential as a function
  of the relative error on $\sin^2\theta_{13}$ for the true value of
  $\delta_{\rm CP}^{\rm TRUE}\,=\,1.45\pi$. The left (right) panel
  represents the results for DUNE (T2HK). The red, green, blue and
  cyan curves delimit the $\theta_{23}$ ``octant-blind'' region
  corresponding to 2, 3, 4 and 5$\sigma$ ($\Delta\chi^2$ = 4, 9, 16, and 25 respectively) confidence.}
\end{figure}

\subsection{ Octant discrimination}

Here we analyse the potential of DUNE and T2HK for excluding the wrong
octant provided the data is generated in the true
octant. Fig.~\ref{fig:13} illustrates the octant sensitivity as a
function of the relative error on $\sin^2\theta_{13}$. The left
(right) panel corresponds to the result for DUNE (T2HK). The colored
curves indicate the sensitivity for discriminating the false octant
from the true one depending on the relative $\sin^2\theta_{13}$ error.
The red, green, blue and cyan correspond to the 2$\sigma$, 3$\sigma$,
4$\sigma$ and 5$\sigma$ confidence level cases, respectively.
NH is assumed as the true hierarchy both in data and theory (note that
the IH case can be considered in exactly the same way). Concerning
theory, we have marginalized over the oscillation parameters
$\theta_{12}$, $\theta_{13}$, $\theta_{23}$, $\delta_{\rm CP}$ and
$\Delta m_{31}^2$ within their allowed range, for a given prior on
$\sin^2\theta_{12}$. One sees from the figure that, depending on the
$\sin^2\theta_{13}$ error, the octant sensitivity increases or
decreases. For example, from the cyan curve for DUNE or T2HK, one sees
that the 1\% error corresponds to 5$\sigma$ sensitivity for
$\sin^2\theta_{23}$(true) $<$ 0.45 and $\sin^2\theta_{23}$(true) $>$
0.58. As the the error increases up to around 6\%, the sensitivity is
gradually lost. So the measurement of the octant of $\theta_{23}$
strongly depends on the relative error of the $\sin^2\theta_{13}$
determination. The octant discrimination sensitivity is slightly
better for T2HK than DUNE due to its high statistics.
\begin{figure}[h!]
\centering
\includegraphics[width=0.5\textwidth]{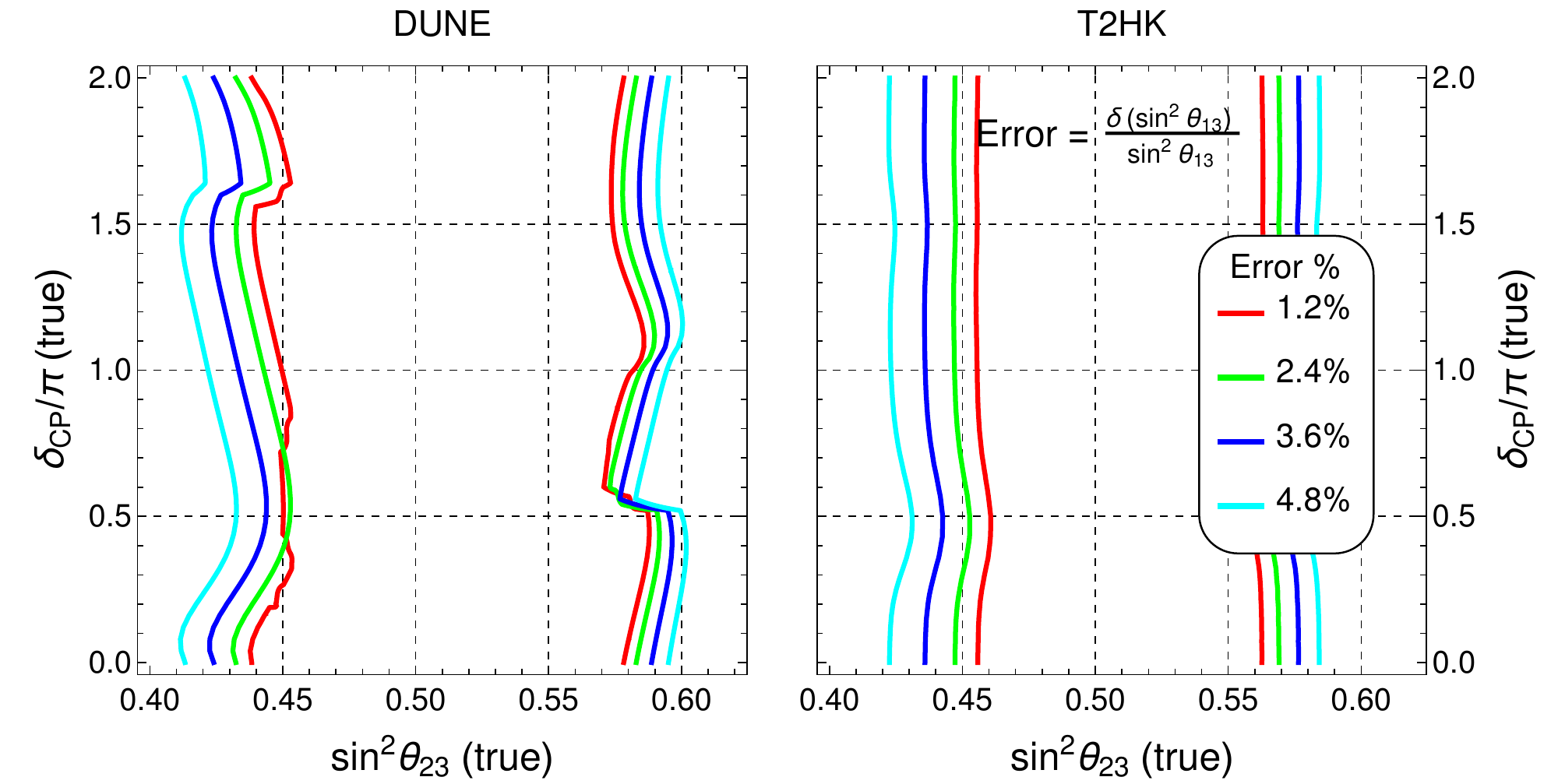}
\caption{\label{fig:13_all} Octant discrimination potential at
  4$\sigma$ ($\Delta\chi^2$ = 19.33) confidence level in the
  $\left[\sin^2\theta_{23},\delta_{\rm CP}\right]$(true) plane.  The
  red, green, blue and cyan curves delimit the ``octant-blind''
  regions corresponding to 1.2\%, 2.4\%, 3.6\% and 4.8\% relative
  errors on $\sin^2\theta_{13}$.}
\end{figure}

In fig.~\ref{fig:13}, we generated the data assuming
$\delta_{\rm CP}$(true) = 1.45$\pi$. So, it is natural to ask what
would be the octant sensitivity over the entire
$\sin^2\theta_{23}$(true) and $\delta_{\rm CP}$(true)
plane. Fig.~\ref{fig:13_all} provides a clear answer to this
question. The simulation details are exactly the same as for
fig.~\ref{fig:13}. This figure neatly summarizes the effect of
the relative $\sin^2\theta_{13}$ error upon the octant sensitivity. We
have assumed a 4$\sigma$ confidence level for the exclusion of the
wrong octant and then varied the various error combinations as
indicated by the different colors.  A band of uncertainty clearly
arises, decreasing the 4$\sigma$ sensitivity range for
$\sin^2\theta_{23}$(true). It is important to notice that our result
is almost independent of $\delta_{\rm CP}$(true). As discussed
earlier, T2HK gives slightly better sensitivity than DUNE.

\subsection{ Other T2HK setups}

\begin{figure}[h!]
\centering
\includegraphics[width=0.5\textwidth]{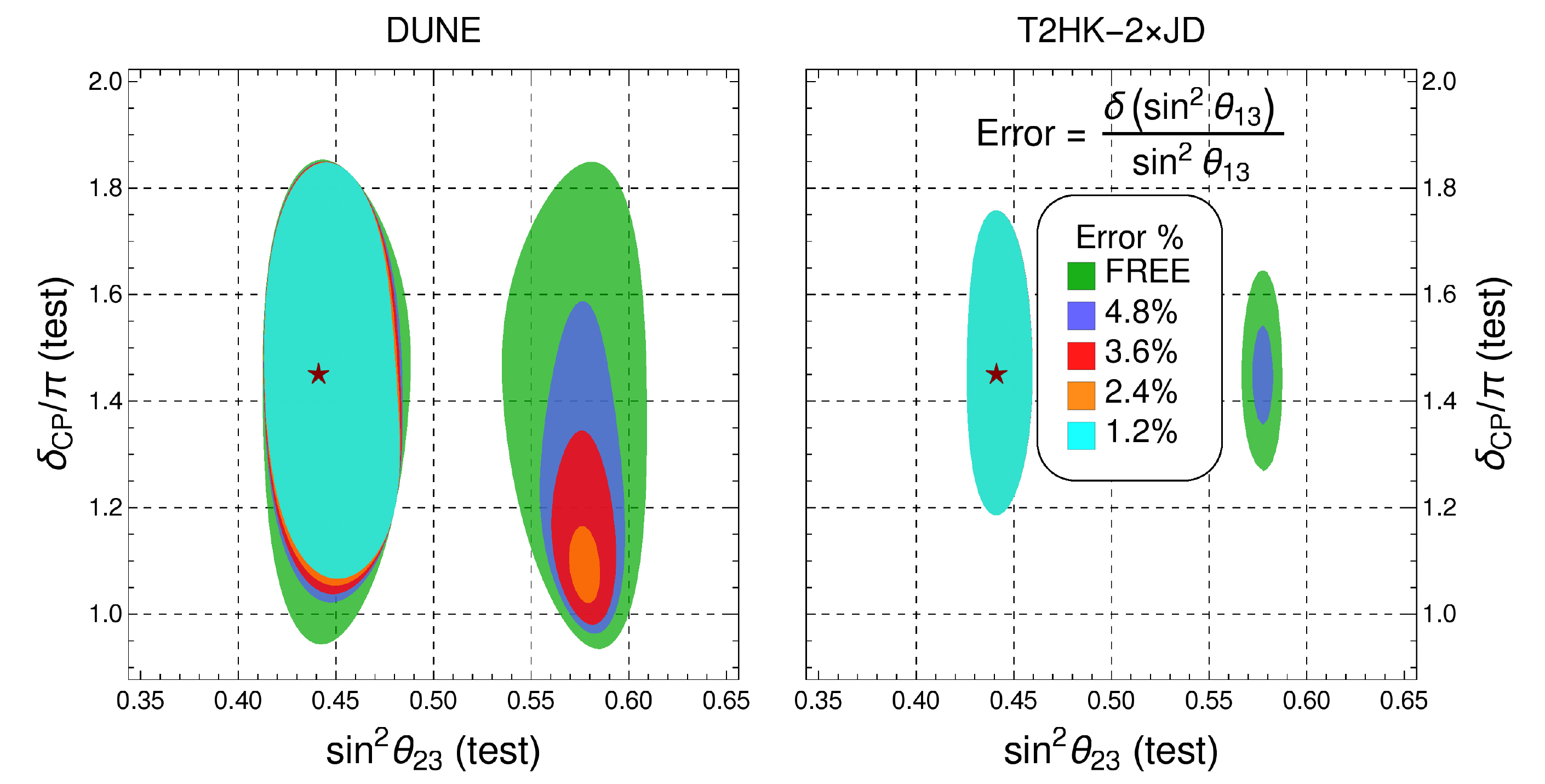}
\caption{\label{fig:14} Measurement of $\theta_{23}$ and
  $\delta_{CP}$ at 4$\sigma$ ($\Delta\chi^2$ = 19.33) confidence. The
  symbol "star" denotes $\sin^2\theta_{23}^{\rm TRUE}$ = 0.441 and
  $\delta_{\rm CP}^{\rm TRUE}$ = 1.45$\pi$. Left (Right) panels
  correspond to DUNE (T2HK). Differently shaded (colored) regions
  correspond to various errors associated with $\sin^2\theta_{13}$. }
\end{figure}

Here we further elaborate upon the T2HK experimental setup. The
    details of the T2HK setup for Fig.~\ref{fig:12} have been already
    described in sec.~\ref{sec:simulation-details}. But for right
    panel of Fig.~\ref{fig:14}, we have considered 295 km of baseline
    and two 187 kton tank as Water Cherenkov far detector placed in
    Japan at an off-axis angle of $2.5^0$.

We call this setup as T2HK-2$\times$JD. A total exposure of (1.3 MW)
$\times$ (10$\times 10^7$ Sec) with a 1:3 $\nu$ and $\bar{\nu}$
running ratio has been assumed. We have assumed an uncorrelated 5\%
(3.5\%) signal normalization, 10\% background normalization error, and
5\% energy calibration error for $\nu$ and $\bar\nu$ appearance (disappearance)
channel. The event numebers and other details have been matched with
\cite{Abe:2016ero}. From these two figures it is clear that the impact
of two setups is not significantly different from each other rather
they are very similar from the perspective of this work. We have also
checked that the result remain also valid for the setup T2HK-JD+KD
following the same reference \cite{Abe:2016ero}, where one detector
with 187 kton fiducial mass is placed in Japan having baseline 295 km
and another detector with 187 kton fiducial mass is placed in Korea
having baseline 1100 km.
 
\section{Conclusions}

Based upon the current global information on neutrino oscillation
parameters we have performed a quantitative analysis of the potential
of upcoming long baseline experiments DUNE and T2HK in resolving the
atmospheric octant ambiguity.
We have found that a precise measurement of the reactor angle
$\theta_{13}$ plays a key role in resolving the octant of the
atmospheric angle $\theta_{23}$ using such future accelerator neutrino
experiments. This highlights the complementarity of reactor and
accelerator-based studies in gaining fundamental information on
neutrino properties.  However, in contrast to the determination
  of the neutrino mass ordering, no matter how good the precision on
  $\theta_{13}$ reached in future studies, there will always be an
  octant blind band in any experiment, close to the limit
  $\theta_{23} \to \pi/4$.

\section*{Acknowledgements}

Work supported by Spanish grants FPA2014-58183-P, Multidark
CSD2009-00064, SEV-2014-0398 (MINECO), PROMETEOII/2014/084
(Generalitat Valenciana). P. S. P. acknowledges the support of
FAPESP/CAPES grant 2014/05133-1, 2015/16809-9 and 2014/19164-6.

\bibliographystyle{apsrev}
\providecommand{\url}[1]{\texttt{#1}}
\providecommand{\urlprefix}{URL}
\providecommand{\eprint}[2][]{\url{#2}}

\end{document}